\documentstyle[12pt,graphicx]{article}
\input colordvi
\def\Journal#1#2#3#4{{#1} {\bf #2}, #3 (#4)}

\def\NPB{{\em Nucl. Phys.} B}
\def\PLB{{\em Phys. Lett.}  B}
\def\PRL{\em Phys. Rev. Lett.}
\def\PRD{{\em Phys. Rev.} D}
\def\ZPC{{\em Z. Phys.} C}

\def\r2{\sqrt 2}
\def\beq{\begin{equation}}
\def\eeq{\end{equation}}
\def\beqn{\begin{eqnarray}}
\def\eeqn{\end{eqnarray}}

\def\sinW2{\sin^2\theta_W}

\def\mz2{M_{z}^2}
\def\c2b{\cos 2\beta}

\def\mz{M_z}

\def\Fq2{F_{2}(q^2)}

\def\beq{\begin{equation}}
\def\eeq{\end{equation}}

\def\gmin2{(g-2)_\mu}

\def\sec2w{sec^2\theta_W}

\def\r2{\sqrt 2}
\def\beq{\begin{equation}}
\def\eeq{\end{equation}}
\def\beqn{\begin{eqnarray}}
\def\eeqn{\end{eqnarray}}

\def\sinW2{\sin^2\theta_W}

\def\mz2{M_{z}^2}
\def\c2b{\cos 2\beta}

\def\mz{M_z}

\def\Fq2{F_{2}(q^2)}
\def\sq2{\sqrt{2}}

\def\sec2w{sec^2\theta_W}

\begin{document}

\begin{titlepage}

\begin{center}
{\large {~ CP VIOLATION IN SUSY, STRINGS AND BRANES}}\\
\vskip 0.5 true cm
\vspace{2cm}
\renewcommand{\thefootnote}
{\fnsymbol{footnote}}
 Tarek Ibrahim$^{a,b}$ and Pran Nath$^{b}$  
\vskip 0.5 true cm
\end{center}

\noindent
{a. Department of  Physics, Faculty of Science,
University of Alexandria,}\\
{ Alexandria, Egypt\footnote{: Permanent address of T.I.}}\\ 
{b. Department of Physics, Northeastern University,
Boston, MA 02115-5000, USA} \\
\vskip 1.0 true cm
\centerline{~ Abstract}
\medskip
A review is given of CP violation in a broad class of models based
on supersymmetry, superstrings and brane models. Such models 
contain typically large CP violating phases which affect a variety of 
supersymmetric phenomenon at low energies and affect search for supersymmetry 
at colliders and in dark matter experiments. 
We focus here on few such phenomena, specifically 
the mixing of the 
CP even and the CP odd  Higgs bosons which can be  induced by loop 
corrections and on CP violation in the muon sector.
Possible signals for the observation of CP violation 
are also discussed.

\end{titlepage}

In this talk we discuss the origin of CP phases in SUSY/string/brane
 models and  the implications of these phases for the
 electric dipole moments of the quarks and the leptons and 
 explore the constraints placed on them by the  current experimental
 limits on the phases. We also investigate the 
 low energy implications of phases for SUSY phenomena at colliders
 and elsewhere. We begin by reviewing  the status of CP in the
 Standard Model. Here  the electro-weak sector of the theory has 
one CP violating phase in the CKM matrix\cite{ckm}. An important constraint
on the CKM matrix is that of unitarity and we display one relation 
that arises  from this constraint\cite{ckm}
\beq
V_{ud}V^*_{ub} + V_{cd}V^*_{cb} + V_{td} V^*_{tb}=0
\eeq		 
The constraint of Eq.(1) can be represented by a unitarity 
triangle whose angles
$\alpha, \beta, \gamma$ are defined by
$\alpha =arg(-{V_{td}V^*_{tb}}/{V_{ud}V^*_{ub}})$,
$\beta =arg(-{V_{cd}V^*_{cb}}/{V_{td}V^*_{tb}})$,
and $\gamma =arg(-{V_{ud}V^*_{ub}}/{V_{cd}V^*_{cb}})$
Currently there are  essentially four direct pieces of evidence for CP
violation seen in nature. Two of these arise  in the neutral Kaon
system in the form of $\epsilon$ and $\epsilon '/\epsilon$ which
are experimentally measured to be
\beq
 \epsilon = (2.28\pm 0.02)\times 10^{-3}, ~~
 \epsilon'/\epsilon =(1.72\pm 0.18)\times 10^{-3} 
 \eeq 
 The third one arises in the neutral system in the
 $B^0_d$ ($\overline{B^0_d}$)$\rightarrow J/\Psi K_s$ decay 
 which gives a direct measurement of $\sin (2\beta)$
 \beqn 
	\sin(2\beta) = (0.75 \pm 0.10 ~~~BaBar; ~~~
    0.99 \pm 0.15 ~~~Belle)
\eeqn	
The fourth piece of evidence in favor of CP violation comes from the
 baryon asymmetry in the universe so that
 \beq
 n_B/n_{\gamma}= (1.5- 6.3)\times 10^{-10}
 \eeq
The results of the first three given by Eqs.(2) and (3) are consistent
with  CP violation given by the Standard Model. Further, there is also
an internal consistency among the results  of the first three, i.e.,
the determination of $\sin(2\beta)$ of Eq.(3)  is compatible
with the indirect constraints on unitarity triangle from $\epsilon$,
 $|V_{ub}/V_{cb}|$, and the mass difference of neutral B mesons 
 $\Delta M_{Bd}$ etc. However, it is well known that an explanation
 of the baryon asymmetry in the universe implies the need of a 
CP violation above and beyond that given by the 
Standard Model\cite{ckm}. 
Further, if future more  precise determinations of the angles
$\alpha, \beta,\gamma$ indicate some breakdown of the unitarity 
triangle that would be one sign of new physics beyond the standard model.
 	Another implication of CP violation is in generation of the
	electric dipole  moments  of elementary fermions. In the
	lepton sector the edms  arise at the multiloop level and 
	are too small to be  observed\cite{hoogeveen}. 
	The results are exhibited 
	in Table 1.
\begin{center} \begin{tabular}{|c|c|c|}
\multicolumn{3}{c}{ Table 1} \\
\hline
  & SM (ecm) & Experiment (ecm)\\
 \hline
 $e$ &  $\leq 10^{-38}$ & $< 4.3 \times 10^{-27}$ \\
 \hline
 \hline
  $\mu$ & $\leq 10^{-35}$   & $ < 1.1\times 10^{-18}$ \\
 \hline
 $\tau$ &  $\leq 10^{-34}$  &  $ < 3.1\times 10^{-16}$\\
\hline
 \hline
\end{tabular}\\
\noindent
\end{center}
The results of Table 1 show that typically the standard model prediction
of the electron edm is more than ten orders  of magnitude smaller than
the current  experimental limit\cite{commins,pdg} 
and the situation for  other leptons
is even worse. There is no conceivable way that in the foreseeable
future experimental accuracy can be improved to the level  needed to
observe the leptonic edms. Thus an observation of a leptonic edm in
the future would be a clear indication of
new physics beyond the Standard Model.
The situation in the quark sector of QCD is more complicated.
Here QCD generates a new CP violation from the term 
$\theta_G\frac{\alpha_s}{8\pi}G\tilde G$ from topological effects.
The effective $\bar\theta =\theta_G+arg (detM_uM_d)+..$ gives a neutron
 edm $d_n\simeq 1.2\times 10^{-16}\bar \theta ecm$ and the current 
 limit ~$ d_n<6.5\times 10^{-26} ~ecm$ implies
 $\bar\theta< 6\times 10^{-10}$. As is well known the smallness of
 $\bar\theta$ poses a problem and how to suppress this contribution
 has been dealt with quite extensively in the literature. 
 The basic idea on how to control it consist of using axions,
 using a massless up quark  or by using a symmetry argument
 to suppress CP violating effects\cite{barrnelson}. Some of the recent 
 variants of these ideas consist of using a 
 gluino-axion model\cite{ma},
 Left-Right models\cite{bdm1}, 
 use of SUSY non-renormalization theorem\cite{schmaltz}
 and gauging away the strong CP problem\cite{ibanez}.

\section{The EDM problem of SUSY, String and Brane\\ Models }
For the rest of the paper we assume that the strong CP problem
has been resolved. However, even with solution of the strong CP problem 
a broad class of models based on supersymmetry, strings and branes
contain a large number of CP violating phases which arise from the
soft SUSY breaking sector of the theory\cite{cpbeyond}. 
These large phases are indeed helpful for baryogenesis, but are 
problematic otherwise in that an order of magnitude calculation
points to violations of the edm constraints.
There are several ways of overcoming these constraints. 
One class  of models consists of just fine tuning the phases to be
small\cite{ellis}. 
Another possibility is that the effect of large  phases on the edm
of the quarks and the leptons are suppressed by making masses 
 in the range of several TeV which would suppress 
 the EDMs\cite{na,dimo}. 
This solution is
contrary to the spirit of SUSY since large masses are contrary to the spirit
of naturalness. A variation of this possibility is to make the
phases in the first two generations  small or vanishing while they are
large in the third generation\cite{chang}. 
While this would produce the desired suppression 
it also constitutes a fine tuning unless small phases are  shown to 
arise in a natural fashion in some string or brane models.
A yet another class of models are those where
 the dangerous phases, i.e. the phases that enter
in the EDMs are small, but otherwise the phases are large\cite{bdm2}.
Finally there is the possibility of internal 
cancellations\cite{incancel,rparity}:
In this mechanism the phases  are typically large but internal cancellations
occur generating a drastic reduction of the edms. Since the smallness
of the edms is  by a cancellation, one expects that the edms should 
be observed by an improvement in experiment be a factor of O(10). 
This possibility has been checked in SUGRA, in MSSM, and in string
and brane models. We note in passing that some of the atomic edms
are also very accurately known\cite{atomic}. However, there are 
significant uncertainties associated with nuclear and
atomic physics effects in the theoretical computations of the 
atomic edms and thus imposition of such constraints has to be done
with care.

\section{CP violation as probe of flavor structure}
   CP violation can act as a probe of the flavor structure of susy
 theories. This can happen if the contribution of the SUSY CP violation 
 to K and B physics is significant. In this context  there are  
 three main scenarios. The first of these is that one has  
negligible contribution from the SUSY phases to K and B physics
and that  all of the CP violation in the K and B physics has 
standard model origin, i.e., arises from  $\delta_{CKM}$.
 In this case SUSY CP phases can still be large, but their 
contribution to the K and B  system is constrained to be small.
This could happen in
a variety of ways such as from mass suppression or from the absence
of new flavor structure in the soft SUSY breaking sector of the theory
beyond what is present in the Yukawas. 
Whatever, the  origin of this suppression in this 
case the K and B systems are not relevant probes of CP violation of
SUSY, string and brane models. Further, one does not find any 
need for a new flavor structure beyond what is present in the 
Yukawas. The second possibility is that there are 
sizable contribution from SUSY phases: Here in addition to the large
SUSY CP phases, a new flavor structure is needed\cite{dine1,masiero1}. 
For example, one needs non-negligible 
flavor changing term in the off diagonal component in the LR mass matrix 
$(\delta_{ij})_{LR}(d)= (m^2_{LR}(d))_{ij}/\tilde m_q^2 $
to get  a significant contribution to  $\epsilon'/\epsilon$.
 Finally, there is a third possibility and that is the extreme viewpoint
that all of the CP phenomena in K and B system arises 
from SUSY phases\cite{frere}
Again in this case a new flavor structure 
is necesary in addition to large phases.
 However, 
there is no compelling reason for this extreme view point.
In any case if one of  the two latter scenarios hold then
 CP violation in the K and B system will act as
a probe of the flavor structure of the theory.

 \section{ Origins of CP violation} 
 We discuss now the possible origins of CP violation is
 SUSY, string and brane models. One possible origin is string  
 compactification. One may call this hard CP violation since this
 type of CP violations can exist even without the breaking of 
 supersymmetry. Now Yukawa couplings which are formed via 
 string compactification will carry this type of CP violation
 and the CKM phase $\delta_{CKM}$ which arises from the Yukawas 
 is therefore  a probe of CP violation arising from string compactification.
 A second source of CP violation is spontaneous symmetry breaking 
 which generates CP phases via the soft breaking parameters.
 Specifically soft  
 SUSY CP phases have origin in spontaneous supersymmetry breaking as
 they arise from moduli fields achieving complex VEV's. In 
 SUGRA/heterotic string models the scale where VEV formation appears
 is the Planck/string scale. However, in gauge mediated
 breaking, or in M theory/brane  models the scale where soft
 CP phases appear could be as low as 10 TeV region.
  Additionally, there is the possibility that new sources of CP 
 violation can occur from spontaneous symmetry breaking at
 the electro-weak scale, e.g., in extensions of MSSM
 with the addition of two Higgs singlets. While 
 spontaneous CP violation does not occur in the Higgs sector of
MSSM, or in NMSSM it can occur in extensions of MSSM
 with the addition of two Higgs singlets\cite{ham1}.
If SUSY contributions to K and B physics turn out to be small,
then one has a rather clean bifurcation, i.e.,
the CP violations in K and B physics are probe of string compactification,
and baryogenesis and other CP phenomena that may be seen in sparticle
decays etc become a probe of spontaneous symmetry breaking.

 We discuss now the question if there is any connection between the
 CKM phase and the SUSY phases. Specifically we want to know if 
 the largeness of the CKM phase has any implication regarding the
 size of the SUSY phases. Now it turns out that CP phases
 arising from the soft parameters are essentially unrelated to
 $\delta_{CKM}$. The reason for this is easily understood since 
 SUSY phases  arise from  spontaneous supersymmetry breaking while
  $\delta_{CKM}$ arises from Yukawa couplings which have their origin
  in the string compactification\cite{cpstrings} and thus largely there
   is not a direct connection
 between the two types of phases. There is, however, one exception to
 this in that the trilinear soft term contains a dependence
 on Yukawas so that\cite{an1} 
\beq
 A_{\alpha\beta\gamma} =F^i\partial_i Y_{\alpha\beta\gamma} +..
 \eeq
 Here we find that large phases of the Yukawas could enter in the
 soft trilinear parameters. Unfortunately this relationship is not
 rigid since large phases can be manufactured for the soft  parameters
 even when the CKM phase is vanishing,
  and conversely the SUSY phases can be  zero even when the CKM
  phase is maximal. 
 For example, in a class of models  $A_0=0$ and thus the
 CP phase of the Yukawas has no influence on the soft parameters.
 However, in a broad class of SUSY/string/brane models large CP phases
 do occur independent of any connection with $\delta_{CKM}$.
 In mSUGRA\cite{msugra,sugra}, $\theta_{\mu}$ and $\alpha_{A_0}$ can be large and similarly 
 in nonminimal sugra models, in heterotic string and brane models the
  phases in general would be large.

\section{CP phases and  SUSY Phenomenon}   
If the CP phases are indeed large they will affect many susy 
phenomena at low energy. First CP phases affect sparticle masses,
decay branching ratios and cross sections\cite{moretti}. Specifically, the
FCNC process $b\rightarrow s+\gamma$, the trileptonic 
signal\cite{trilep}
and collider phenomenology is affected. CP effects show up in 
K and B physics and quantities such as $\epsilon'/\epsilon$ are
affected\cite{masiero1}. Quite interestingly the supersymmetriic 
contribution to
 $g_{\mu}-2$ is strongly affected\cite{ing}. The CP phases enter in the
 analysis of Higgs physics leading to mixing of CP even and CP
 odd neutral sectors\cite{pilaftsis,carena,inhiggs1,inhiggs2}. 
 These mixings will have many interesting
 features including new signals at 
 colliders\cite{barger,zerwas,voloshin}. Similar phenomena will
 occur for soft gaugino masses in experiments at 
 colliders\cite{mrenna}  and in phenomenology of 
 sleptons\cite{bartl}. 
 Other phenomena 
 which are sensitive to CP violation are the analyses of neutralino
 relic density\cite{cin},
 proton decay\cite{cpproton} and 
 baryogenesis\cite{baryogenesis}. Additionally, there are  a 
 variety of other phenomena not yet investigated where the 
 CP phases are likely to enter strongly. One now must ask 
  how experiment will determine the phases. In general this is 
  a more complicated question than what one might imagine.
  The reason is that while for mSUGRA case one has two phases 
  which one can choose to be the phase of $\mu$  and the phase 
  of $A_0$, for the more general soft SUSY breaking scenarios
  there are in general a large number of independent phases
  which enter in various combinations in susy phenomena. 
  In Table 2 we  exhibit some examples of the processes and list
  the combination of the phases that enter in that process.
  One finds that in general a susy process will contain several
  combination of susy phases and thus one will need  measurement 
  of several process to pin down the phases. Some examples of the
  combinations of phases that enter in SUSY phenomena are  given
  in Table 2.

\begin{center} \begin{tabular}{|c|c|}
\multicolumn{2}{c}{Table 2: Examples of CP phases in  SUSY phenomena } \\
\hline
SUSY Quantity  & Combinations of CP violating phases \\
\hline
$m_{\tilde W}$ ($m_{\chi_i}$)  &  $\xi_2+\theta_{\mu}$ 
($\xi_2+\theta_{\mu}$, $\xi_1+\theta_{\mu}$) \\
\hline
$b\rightarrow s+\gamma$ & $\alpha_{A_t}+\theta_{\mu}$,  
$\xi_2+\theta_{\mu}$, $\xi_3+\theta_{\mu}$,
  $\xi_1+\theta_{\mu}$\\
\hline
$\tilde W\rightarrow q_1\bar q_2 + \chi_1$,..& $\xi_2+\theta_{\mu}$,
$\alpha_{A_{q1}}+\theta_{\mu}$,$\alpha_{A_{q2}}+\theta_{\mu}$,
$\xi_1+\theta_{\mu}$,.\\
\hline
$\tilde g\rightarrow q\bar q + \chi_1$,..& $\xi_2+\theta_{\mu}$,
$\alpha_{A_{q}}+\theta_{\mu}$, $\xi_2+\theta_{\mu}$, $\xi_1+\theta_{\mu}$,.\\
\hline
$g_{\mu}-2$ & $\xi_2+\theta_{\mu}$, $\xi_1+\theta_{\mu}$,
$\alpha_{A_{\mu}}+\theta_{\mu}$\\
\hline
$m_{H_i}$(small $\tan\beta$) & $\alpha_{A_t}+\theta_{\mu}$  \\
\hline 
 $m_{H_i}$(large $\tan\beta$) 
  & $\alpha_{A_t}+\theta_{\mu}$, $\alpha_{A_b}+\theta_{\mu}$,
   $\xi_2+\theta_{\mu}$,  $\xi_1+\theta_{\mu}$\\
   \hline
  $Z^*\rightarrow Z+H_i$ & $\alpha_{A_t}+\theta_{\mu}$, $\alpha_{A_b}+\theta_{\mu}$,
   $\xi_2+\theta_{\mu}$,  $\xi_1+\theta_{\mu}$\\
 \hline  
 $d_e$ ($d_{\mu}$)  &   $\xi_2+\theta_{\mu}$, $\xi_1+\theta_{\mu}$,
 $\alpha_{A_e}+\theta_{\mu}$($\alpha_{A_e}+\theta_{\mu}$ )\\
 \hline
 $d_n$ & $\xi_3+\theta_{\mu}$,  
 $\xi_2+\theta_{\mu}$, $\xi_1+\theta_{\mu}$,
 $\alpha_{A_{ui}}+\theta_{\mu}$,$\alpha_{A_{di}}+\theta_{\mu}$ \\
 \hline
\end{tabular}
\end{center} 
Table Caption:  $\theta_{\mu}$ is the phase of the Higgs mixing parameter
$\mu$, 
$\xi_i$ is the phase of gaugino mass $\tilde m_i$ (i=1,2,3) and
$\alpha_{A_q}$ is the phase of trilinear coupling $A_q$.     


\section{Effects of CP violation in the Higgs sector}
One of the interesting phenomenon of soft SUSY CP violating phases
 in that they induce a CP violation in the Higgs sector at the one
 loop level\cite{pilaftsis,inhiggs1,inhiggs2}. To account for the
 induced CP violation in the Higgs sector one can parametrize the 
Higgs fields so that 

\beqn
(H_1)= \left(\matrix{H_1^0\cr
 H_1^-}\right)
 =\frac{1}{\sqrt 2} 
\left(\matrix{v_1+\phi_1+i\psi_1\cr
             H_1^-}\right) \nonumber\\
(H_2)= \left(\matrix{H_2^+\cr
             H_2^0}\right)
=\frac{e^{i\theta_H}}{\sqrt 2} \left(\matrix{H_2^+ \cr
             v_2+\phi_2+i\psi_2}\right)
\eeqn
In the basis   $\{ \phi_1,\phi_2,\psi_{1D}, \psi_{2D}\}$ defined by
\beqn
\psi_{1D}=\sin\beta \psi_1+ \cos\beta \psi_2\nonumber\\
\psi_{2D}=-\cos\beta \psi_1+\sin\beta \psi_2 
\eeqn
$\psi_{2D}$ decouples and the remaining $3\times 3$ matrix is 

\beq
M^2_{Higgs}=
\left(\matrix{M_Z^2c_{\beta}^2+M_A^2s_{\beta}^2+\Delta_{11} &
-(M_Z^2+M_A^2)s_{\beta}c_{\beta}+\Delta_{12} &\Delta_{13}\cr
-(M_Z^2+M_A^2)s_{\beta}c_{\beta}+\Delta_{12} &
M_Z^2s_{\beta}^2+M_A^2c_{\beta}^2+\Delta_{22} & \Delta_{23} \cr
\Delta_{13} & \Delta_{23} &(M_A^2+\Delta_{33})}\right) 
\eeq
The analysis for the cases of $t-\tilde t$\cite{pilaftsis,inhiggs1,inhiggs2} 
and 
$W,H^+,\tilde W$\cite{inhiggs1} exchanges is straightforward
and can be carried out analytically since diagonalization of only
$2\times 2$ matrices are involved.
The analysis of $Z, A, H^0, \chi^0$ exchange is more involved  
and  requires a calculus of eigenvalues\cite{anloop,carena,inhiggs2}.
 We review here briefly this technique. The loop corrections
 to the Higgs $(mass)^2$ matrix are in general given by 

\beq
\Delta M_{ab}^2=
\frac{1}{32\pi^2}
\sum_i(\frac{\partial \lambda^2_i}{\partial \Phi_a}\frac{\partial \lambda^2_i}
{\partial\Phi_b}
log\frac{\lambda^2_i}{Q^2}+\lambda^2_i\frac{\partial^2 \lambda^2_i}
{\partial \Phi_a\partial \Phi_b}
log\frac{\lambda^2_i}{eQ^2})_0
\eeq
The computation of the derivatives for the neutralino mass matrix
requires special attention. 
We show that  although the eigen values of an $n\times n$ $(mass)^2$
matrix  cannot be analytically computed one can compute analytically
the derivatives of the eigen values in terms of the co-efficients
of the polynomial that defines the eigen value equation.
Thus consider an nth order eigen value equation
\beq
F(\lambda)=Det(M^{\dagger}M -\lambda I)=
\lambda^n+c^{(n-1)}\lambda^{n-1}+c^{(n-2)}\lambda^{n-2}+..
+c^{(1)}\lambda  +c^{(0)}=0
\eeq
The co-efficients are explicit functions of the background
fields 
$\Phi_{\alpha}=\{\phi_1,\phi_2,\psi_1,\psi_2\} $.
while the eigen values are implicit functions of the
background fields through the satisfaction of the eigen value equation.
One can now establish that
\beq
\frac{\partial\lambda_{i}}{\partial\Phi_{\alpha}}
=-(\frac{D_{\alpha}F}{D_{\lambda}F})_{\lambda=\lambda_{i}}
\eeq
where  $D_{\lambda}$ differentiates the $\lambda$ dependence in
$F$,
$D_{\lambda}F(\lambda)={dF}/{d\lambda}$ 
and $D_{\alpha}$ differentiates only the co-efficients,
i.e., $D_{\alpha}F=c^{(n-1)}_{\alpha}\lambda^{(n-1)}+
c^{(n-2)}_{\alpha}\lambda^{(n-2)}+..+c^{(1)}_{\alpha}\lambda 
+c^{(0)}_{\alpha}$, and $[D_{\alpha},D_{\lambda}]=0 $.
These equations provide us with a technique for 
analyzing cases where the analytic solutions to the eigen values
are not available. It is well known  that the 
$t-\tilde t$ exchange generates a large CP even-CP odd higgs
 mixing\cite{pilaftsis,inhiggs1}. However, 
significant contributions can arise from the 
$\tilde W-W-H^+$ exchanges especially for large  
$\tan\beta$\cite{inhiggs1}.
 Specifically, for $\tan\beta \geq 30$ the chargino 
contributions can dominate the stop contribution. 
Significant contributions also arise from the 
$Z, A, H^0, \chi^0$ exchanges\cite{inhiggs2}. 
The contributions from this sector
are comparable to the contributions from the stop and
chargino sectors for values of $\tan\beta \geq 5$.
If large CP phases exist, then collider experiments will 
provide signals such as three peaks in $Z^*\rightarrow Z+H_i$ and
modified rates of $h\rightarrow b\bar b$. Further, signals may 
emerge from decays of the neutral and charged Higgs bosons of the
type $H^0\rightarrow  \chi^{\pm}_i\chi^{\mp}_j$ and 
$H^{\pm}\rightarrow  \chi^{0}_i\chi^{\pm}_j$\cite{drees}.
An interesting observation in made in the analysis 
of Ref.\cite{inhiggs3} in that if 
 a mixing effect is observed experimentally
then among the three possibilities, i.e., the fine tuning,
the heavy sparticle spectrum, and the cancellation mechanism, 
it is only the cancellation mechanism that can survive under the 
naturalness constraint\cite{inhiggs3,ccn}.

\section{ CP violation in the muon system}
The supersymmetric correction to the muon anomalous magnetic moment
$a_{\mu}^{SUSY}=(g_{\mu}-2)/2$\cite{yuan} 
is a sensitive function of the phases and shows
a rapid variation with the $\mu$ phase 
and the SU(2) phase $\xi_2$\cite{ing}.
As a consequence of the phases the chargino contribution need not be
much larger than the neutralino contribution  to 
$a_{\mu}^{SUSY}$ as is usually the case\cite{icn}. If an 
$a_{\mu}^{SUSY}\geq 10^{-10}$  emerges at BNL then
this limit will significantly constrain CP phases\cite{icn,arnowittedm}.
Further, it is possible to generate models  with low sparticle
spectra satisfying BNL and EDM constraints\cite{icn}.
There is a recent proposal to  probe $d_{\mu}$ with a sensitivity of
 $d_{\mu}\sim O(10^{-24})ecm$\cite{yanni}.
In most theoretical models the charge lepton edms scale, i.e.,
\beq
\frac{d_{\mu}}{d_e}\simeq \frac{m_{\mu}}{m_e}
\eeq
 $d_e< 4.3\times 10^{-27}ecm$ implies 
$d_{\mu}<10^{-25}ecm$
 below the sensitivity of the proposed BNL experiment.
Large muon edms can be gotten only by the breakdown of scaling, e.g.,
in the two higgs doublet model\cite{ng}, in Left-Right models\cite{bdmedm}, 
and in models with non-universalities in the slepton 
sector\cite{inmuedm,feng} where 
$\alpha_{A_{\mu}}\neq \alpha_{A_e}$, $|A_{\mu}|\neq |A_e|$.

~
\section{Conclusions}
The discussion given here shows that CP violation is an important 
probe of susy/string/brane models. It is most likely that there are
more than one origin of CP violation. One of these is string 
compactification, and another, spontaneous symmetry breaking. 
Thus CP violation is a probe of string compactification as well as 
of symmetry breaking. Further, CP violation could also be a probe of
the flavor structure of  supersymmetric models if the SUSY contributions
to K and B physics are significant. One also finds that the 
edms if observed could provide a further probe of the flavor structure
of supersymmetric theories. In this context the proposed BNL
experiment to measure the muon edm at the sensitivity of $10^{-24}ecm$ is 
important as a probe of the flavor structure of susy/string/brane models.  
Further, collider experiments have the potential to tell a lot about CP
violation specifically regarding the existence of CP beyond the K and B
systems. This will occur via analyses of sparticle masses and decays,
production cross sections, and via possible observation of CP even-CP odd 
mixing in the neutral  Higgs system. 

\section{Acknowledgments}
This research was supported in part by NSF grant PHY-9901057.

%


\end{document}